# Leveraging Quadratic Polynomials in Python for Advanced Data Analysis

**[revision 1]**

Rostyslav Sipakov, Ph.D.,[1*] Olena Voloshkina, D.Sc.,[1] Anastasiia Kovalova, Ph.D.,[1]

[1]Kyiv National University of Construction and Architecture, Department of Environmental Protection and Labor Protection Technologies, Povitroflotskyi Ave, 31, Kyiv, Ukraine, 03037.

[*]Correspondence concerning this article should be addressed to Dr. Sipakov, e-mail: sipakov.rv@knuba.edu.ua

## Abstract

This research explores the application of quadratic polynomials in Python for advanced data analysis. The study demonstrates how quadratic models can effectively capture nonlinear relationships in complex datasets by leveraging Python libraries such as NumPy, Matplotlib, scikit-learn, and Pandas. The methodology involves fitting quadratic polynomials to the data using least-squares regression and evaluating the model fit using the coefficient of determination (R-squared). The results highlight the strong performance of the quadratic polynomial fit, as evidenced by high R-squared values, indicating the model's ability to explain a substantial proportion of the data variability. Comparisons with linear and cubic models further underscore the quadratic model's balance between simplicity and precision for many practical applications. The study also acknowledges the limitations of quadratic polynomials and proposes future research directions to enhance their accuracy and efficiency for diverse data analysis tasks. This research bridges the gap between theoretical concepts and practical implementation, providing an accessible Python-based tool for leveraging quadratic polynomials in data analysis.

**Keywords:** python, quadratic polynomials, analyzing data,

## 1. Introduction

In exploring the quadratic polynomials used in various applications in Python for data analysis, we found significant contributions across various domains, exemplifying their utility and versatility. Quadratic polynomials have been found to be extensively valuable in fields such as physics, fluid mechanics, and chemical reaction modeling due to their effectiveness in capturing non-linearities.

The primary advantage of quadratic polynomials lies in their ability to capture curvature, allowing them to model complex patterns beyond the reach of traditional linear models. This capacity to represent curved relationships is crucial for accurately depicting trends in data, particularly when there are significant changes in the rate of increase or decrease. Despite their ability to handle non-linear relationships, quadratic polynomials remain relatively simple and interpretable (Fuchs et al., 2009). Additionally, they are computationally efficient compared to higher-degree polynomials or more complex non-linear models. For example, theoretical physics has employed actions quadratic in curvature and polynomial in variables to effectively describe observer spaces (Koivisto et al., 2019). In computer science, quadratic polynomial interpolation has been applied in 3D path planning to derive continuous paths for each axis (Chang & Huh, 2015). When compared to linear and cubic polynomials, quadratic polynomials are particularly useful in scenarios where relationships are non-linear but not overly complex to necessitate higher-order polynomials. Linear models struggle to capture curvature and non-linear patterns, while cubic polynomials may introduce unnecessary complexity where a quadratic model would suffice (Gibert et al., 2000).

As mentioned above, the use of quadratic polynomials in modeling has proven beneficial in various applications. Aladesanmi et al. (2021) illustrated the adaptability of quadratic polynomials in material science and applied these models to understand the wear rate and hardness of Ti and TiB2 nanocomposites. Their research findings, indicating a better fit for the quadratic model with an adjusted R-squared value of 0.8883, underscores the utility of quadratic polynomials in material science research. In epidemiology, Yadav (2020) leveraged quadratic polynomial regression models to analyze the COVID-19 epidemic in India, demonstrating its effectiveness in epidemic forecasting. This example reflects the predictive power of mathematical models and their crucial role in public health planning and responses (Yadav, 2020 ). In the context of urban development and assessment of geotechnical conditions, the incorporation of Python for data analysis, particularly through quadratic polynomials, can significantly enhance the understanding and monitoring of complex ground conditions (Kaliukh et al., 2022). In the context of quadratic polynomial regression, it is essential to note that quadratic polynomial step regression is an advanced tool capable of utilizing orthogonal experimental data to build a regression model, while avoiding instability in the regression coefficients owing to the multicollinearity of the variables (Wang et al., 2014). This highlights the potential of quadratic polynomials for handling complex data relationships and providing accurate regression models. Gong and Zhang (2021) developed a polynomial regression model to predict Python usage trends. Their model, which demonstrated high accuracy with a training set score of 0.912862 and a test set score of 0.886600, highlighted the effectiveness of quadratic polynomials in forecasting software usage patterns (Gong and Zhang, 2021).

Python, a high-level programming language, provides an ideal environment for the rapid prototyping of data analytic tools and includes powerful tools for visualization, data sharing, and statistical analysis, such as Matplotlib, iPython, NumPy, and SciPy (Alexander et al., 2017). Gong and Zhang (2021) presented a compelling application for predicting Python usage trends and demonstrated a robust model fit with practical implications for software analytics. In summary, Python, with its extensive libraries and capabilities for rapid prototyping, visualization, and scientific computation, provides a robust platform for leveraging quadratic polynomials in advanced data analysis tasks.

## 2. Methods

### 2.1 Design and development environment

In this study, we focused on applying quadratic polynomials in Python for data analysis, highlighting the importance of these mathematical expressions in modeling and interpreting complex datasets using the following key concepts:

- Quadratic polynomials: Defined by the general form $ax2 + bx + c$ , where $(a), (b), (c)$ , are coefficients. These polynomials are essential for capturing curvature in datasets indicative of various natural and human-made phenomena.

- Python libraries: NumPy is open source and is available at https://numpy.org, were used for numerical computations, and Matplotlib also is open source and is available at https://matplotlib.org), was used to plot the data and polynomial curves, showing how these tools were integrated for data analysis. Additionally, we applied the following open source libraries: "scikit-learn" is a popular machine learning library in Python that offers a plethora of features that make it a preferred choice for machine learning practitioners, available at https://scikit-learn.org, and Pandas, which is a powerful tool in Python for data manipulation and analysis, and available at https://pandas.pydata.org.
- Regression analysis: Explains how quadratic polynomials can be fitted to data points to model relationships within the data, emphasizing practical applications through Python coding examples.
- Coefficient of determination (R-squared): Discuss the computation and interpretation of R-squared to measure how well the polynomial model fits the data.

A quadratic polynomial is an algebraic equation of the second degree, which includes a term raised to a power of two (squared). The general form of a quadratic polynomial is $y = ax2 + bx + c$, where $(y)$ is the dependent variable; $(x)$ is the independent variable; and $(a), (b), (c)$ are the coefficients of the polynomial estimated by the regression model. The quadratic term $(ax2)$ allows the model to capture the curvature in the data, which is indicative of acceleration increases or decreases that are common in many natural phenomena.

Some key features of quadratic polynomials are that they have two terms with a variable $(x)$ - one is $(x)$ - squared, and the other is $(x)$ to the first power. The $(x2)$ term has a non-zero coefficient $(a)$. This makes it a quadratic polynomial rather than a linear polynomial. When plotted, quadratic polynomials form a parabolic shape rather than a straight line. The quadratic polynomials have up to two distinct real roots for the equation $x2 + bx + c = 0$. These solutions were obtained by factoring or by using a quadratic formula. Examples of quadratic polynomials include the vertex form $y = a(x - h)2 + k$, and the standard form $y = ax2 + bx + c$. A quadratic polynomial has a squared, linear, and constant term, graphs as a parabola, and two roots at most.

Understanding their structures allows many mathematical and real-world problems to be solved. We provide an example of the Python script below, which employs a quadratic polynomial fitting technique–a method used in regression analysis to model the relationship between a dependent variable and one or more independent variables. In this case, the independent variable is time (represented in months), and the dependent variable is the metric of interest (such as pollution levels).

After fitting the quadratic polynomial to the data, the script generated a smooth-fitted curve that represented the estimated values of the dependent variable across a range of independent variables. This curve helps to visualize the overall trend and any potential seasonal patterns or anomalies in the dataset.

The coefficient of determination, commonly known as R-squared (R2), was then calculated to quantify the goodness of fit of the polynomial model. It is a statistical measure that indicates the proportion of variance in the dependent variable that is predictable from independent variable(s). An (R2) value of one (1) indicated a perfect fit, indicating that the model explained all the data variability around its mean. In contrast, an (R2) value closer to zero (0) indicates that the model fails to accurately model the data.

For more in-depth information on quadratic polynomial fitting and calculation of the coefficient of determination, the following sources (Norman R. Draper and Harry Smith, 2014; Douglas et al., 2021) provide a comprehensive overview of seminal works on regression analysis and detailed explanations of various regression techniques, including quadratic polynomial fitting and interpretation (R2) .

Next, we applied quadratic polynomial fitting and R-squared in Python for the data analysis.

In this case, the Python script exemplifies the application of regression analysis using the NumPy, Matplotlib, scikit-learn, and Pandas libraries to model and visualize trends in time-series data. A core component of this analysis is the fitting of a quadratic polynomial to the data, grounded in the principles of statistical learning.

The schematic block diagram on Figure 1 shows the framework of the explanted application in this research: implementing quadratic polynomials in Python for advanced data analysis.

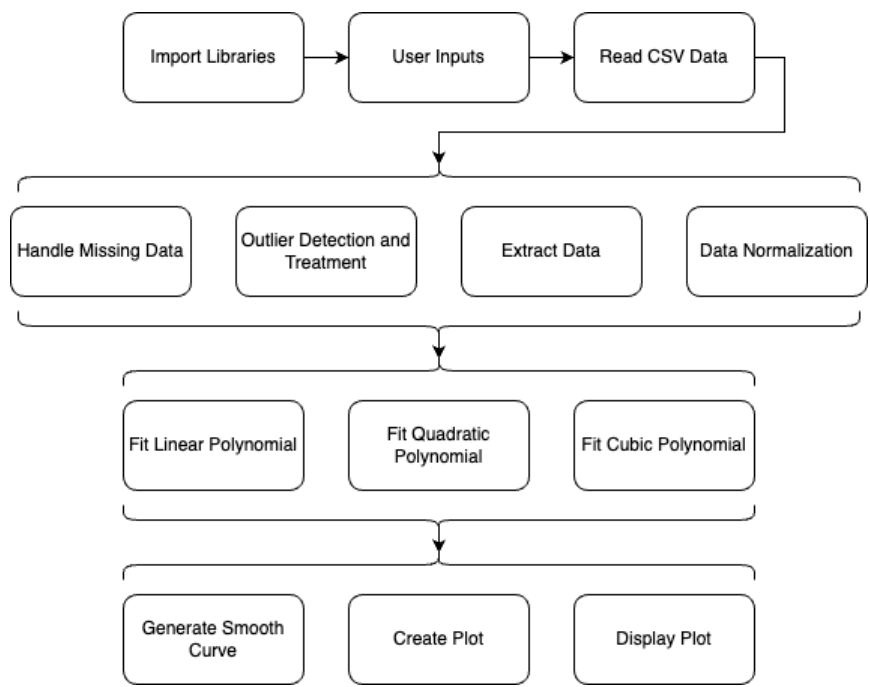

**Figure 1. The schematic block diagram of the implementing quadratic polynomials in Python for advanced data analysis**

## 2.2 Data Preprocessing

### 2.2.1. Handling missing data

Before fitting the quadratic model, it is crucial to address any missing values in the dataset to ensure the accuracy and reliability of the model. Common techniques for handling missing data include:

a. Removal of Missing Values

   This involves deleting rows or columns that contain missing values. This method is straightforward but may lead to loss of valuable information if the dataset has many missing entries. The Python-formatted version of the code snippet for this technique is similar to:

```python
# Removing rows with missing values
data_cleaned = data.dropna()
```

    b. Imputation

       This technique fills in missing values with plausible data. Common imputation methods include:

       Mean or Median Imputation - Replaces missing values with the mean or median of the column. The Python-formatted version of the code snippet for this technique is similar to:

```python
# Mean or Media Imputation
data['column_name'].fillna(data['column_name'].mean(), inplace=True)
```

       Forward or Backward Fill - Uses previous or next observation to fill missing values. The Python-formatted version of the code snippet for this technique is similar to:

```python
# Forward or Backward Fill
data.fillna(method='ffill', inplace=True)
```

In the context of machine learning, the exclusion of rows containing missing values without the application of imputation techniques can result in substantial data attrition, particularly in datasets with a high incidence of missing entries. This practice can diminish the statistical power of the analysis and introduce bias into the results. Imputation mitigates this bias by populating missing values with plausible data points, thus enabling the model to utilize the maximum amount of available data. Models that are trained on complete datasets, including imputed values, exhibit greater consistency and robustness compared to those trained on datasets with omitted missing values (Liu et al., 2020; Dubey & Rasool, 2021).

For numerical data, simple imputation methods, such as substituting missing values with the mean or median, are both effective and straightforward. This approach operates under the assumption that the data is missing at random, thereby not introducing significant bias. In the context of more complex datasets, advanced imputation techniques, including K-Nearest Neighbors (KNN), regression imputation, or multiple imputation, can yield more accurate estimations of missing values (Suh et al., 2023; Karrar, 2022). It is important to note that numerous machine learning algorithms are incapable of processing missing data and will either fail or generate errors in the presence of such data (Fleck et al, 2020). Imputation ensures the completeness and compatibility of the dataset with these algorithms.

Appropriate imputation techniques are crucial for preserving the inherent relationships and variability within the data, which in turn enhances the accuracy and reliability of the resulting models.

Let's consider a situation where we are analyzing the impact of outdoor air pollution at a highway intersection on different days of the week. Depending on various factors, data may not always be recorded daily. Thus, when analyzing a dataset for an entire calendar year, simply removing rows with missing data would not provide a complete assessment of the full year. Imputation allows us to fill these gaps, maintaining a complete and continuous dataset for analysis. For real practice cases, for example, the incidence of secondary air pollution by formaldehyde, arising from photochemical reactions in urban environments, is exhibiting an increasing trend (Voloshkina et al., 2019). To accurately assess non-carcinogenic and carcinogenic risks and thereby mitigate disease incidence in the population, it is imperative to utilize the most comprehensive datasets of air pollution parameters (Sipakov et al., 2018). However, obtaining such extensive datasets is often not technically feasible. In such scenarios, the aforementioned imputation can play a pivotal role in addressing these data deficiencies.

Additionally, we would like to note that in the context of the dataset and its analysis, the user needs to determine what is most appropriate for the given case: the technique of removing rows with missing values or imputation based on a number of factors. With a small amount of missing data, up to five (5) percent, removing such rows from the dataset will not significantly impact the analysis results. However, with a large amount of missing data, more than 5 percent, simply deleting rows can lead to a significant loss of information, making imputation the most appropriate option. If the data in the dataset under consideration is missing randomly, filling in using the mean or median value may be a more effective method. If there is a pattern in the missing data (for example, specific days of the week), more complex filling methods specific to the particular domain will be required. However, it is important to pay significant attention to the specific application of the dataset. For example, if we are analyzing daily air pollution indicators collected every day, data from Monday to Thursday from 8.00 AM to 5.00 PM in urban areas will likely be similar to each other under stable weather conditions. But they can differ significantly with changes in weather conditions, day of the week (Monday vs. Saturday), and time of day (day vs. night), particularly in terms of traffic congestion on highways, for instance.

### *2.2.2. Outlier Detection and Treatment*

While collecting data from air pollution sensor recorders, we encountered anomalous readings that were either significantly lower or higher than expected. This could be due to temporary equipment malfunctions or sensor contamination. Therefore, in addition to removing empty rows from the dataset, we must process rows containing outlier data.

Outliers are data points that differ significantly from other observations. They can significantly affect the performance of the quadratic model and distort its results, making it essential to detect and handle them properly (Yerlikaya-Özkurt et al., 2016).

Various methods have been proposed to detect outliers, such as nonparametric approaches and robust regression techniques (Fan et al., 2006; Toshiaki et al., 2021). These methods aim to identify and remove outliers effectively to ensure the robustness of the analysis. Outlier detection often involves comparing the local density of data points to that of their neighbors, highlighting the importance of considering the context of neighboring points (Latecki et al., 2007).

Our study will examine two main methods to detect outliers: statistical and data normalization or scaling. Normalization or scaling ensures that different features contribute equally to the analysis, which is crucial when features have varying units or scales.

   a. **Z-Score Method**

   The Z-score method is a commonly used statistical technique for identifying outliers in data by measuring how many standard deviations a data point is from the mean.

Data points with Z-scores greater than 3 or less than -3 are typically considered outliers (Yaro, 2024). To enhance the Z-score method, a modified Z-score based on the median can be utilized to improve robustness, especially when dealing with data values that significantly differ from the mean (ÇILGIN et al., 2023). This modification helps mitigate the sensitivity of traditional Z-scores to extreme data values. Despite the fact that this method has a preference over more recent techniques due to its simplicity and reliability, it is essential to recognize that the Z-score method may be vulnerable to masking effects, where the presence of outliers can distort the sample mean and variance, potentially concealing other outliers within the data. Therefore, while the Z-score method holds significant value, it is imperative to be aware of its limitations and to consider employing supplementary techniques for a more comprehensive outlier detection strategy. The Python-formatted version of the code snippet for this technique is similar to:

```python
# Z-Score Method

z_scores = stats.zscore(data['column_name'])
abs_z_scores = np.abs(z_scores)
filtered_entries = (abs_z_scores < 3)
data_cleaned = data[filtered_entries]
```

b. **The Interquartile Range (IQR)**

The Interquartile Range (IQR) method is a statistical technique used to detect outliers in a dataset. It involves calculating the IQR, which is the difference between the 75th percentile (Q3) and the 25th percentile (Q1) of the data. Data points that fall outside the range of Q1 - 1.5 * IQR and Q3 + 1.5 * IQR are considered outliers. This is a common rule of thumb known as the 1.5 IQR rule, which is widely used in various fields such as data science, environmental monitoring, and outlier detection in different applications. For example, in the assessment of ICESat-2 laser altimeter data for water-level measurement, the IQR method was applied to detect outliers (Cui et al., 2020). The Python-formatted version of the code snippet for this technique is similar to:

```python
# The Interquartile Range (IQR)

Q1 = data['column_name'].quantile(0.25)
Q3 = data['column_name'].quantile(0.75)
IQR = Q3 - Q1
filter = (data['column_name'] >= Q1 - 1.5 * IQR) & (data['column_name'] <= Q3 + 1.5 * IQR)
data_cleaned = data.loc[filter]
```

c. **Min-Max Scaling**

Min-max scaling, also known as min-max normalization, is a data normalization technique that transforms features so that they fall within a specific range, typically between zero (0) and one (1) (Ampomah et al., 2021). This method entails rescaling data to ensure it lies within the interval of zero to one, facilitating a standardized comparison of values both before and subsequent to processing. Min-max normalization is particularly critical in data preprocessing as it preserves the relationships among the original data points while ensuring they reside within a consistent range. The min-max normalization technique effectively scales the data to conform to these specified boundaries by determining the new minimum and maximum values. Researchers have highlighted the significance of min-max normalization in various fields, such as machine learning, where it is considered one of the most common tools for data normalization (Gertz et al., 2019). The Python-formatted version of the code snippet for this technique is similar to:

```python
# Min-Max Scaling

scaler = MinMaxScaler()
data_scaled = scaler.fit_transform(data)
```

    d. **Standardization**

    Standardization, a common normalization technique, involves centering data around the mean and scaling it to have a unit standard deviation. This process is widely used in various fields. For instance, in image analysis, normalizing input data with mini-batch statistics enforces elements in feature maps to have a zero mean and unit standard deviation (Wu et al., 2020). The Python-formatted version of the code snippet for this technique is similar to:

```python
# Standardization

scaler = StandardScaler()
data_standardized = scaler.fit_transform(data)
```

## 2.3 Fitting the Quadratic Model

In Python, this was achieved using the `Polynomial.fit` method from the NumPy library, which computes the least-squares fit of a polynomial of a specified degree to the given data. The snippet calculates the optimal values for coefficients $(a), (b), (c)$ that minimizes the sum of the squared differences between the observed values and values predicted by the polynomial, thereby effectively "fitting" the curve to the data, which Python-formatted version of code snippet similar to:

```python
# Fit the quadratic polynomial
coefs = Polynomial.fit(months, values, 2).convert().coef
```

With the fitted polynomial, our script generates a curve across a continuum of points within the data range, which was visualized using Matplotlib's plotting function and the Python-formatted version of the code snippet similar to:

```python
# Generate a smooth curve by evaluating the polynomial at many points

x = np.linspace(months.min(), months.max(), 200)
y = coefs[0] + coefs[1] * x + coefs[2] * x**2

# Plot the data and the fitted curve
plt.plot(x, y, color='purple', label='Fitted curve')
```

The coefficient of determination, R2, was subsequently computed to assess the fit quality. Python was used to compare the variance of the residuals (the differences between the observed and predicted values) with the total variance of the data, and the corresponding code snippet is similar to:

```python
# Calculate R-squared value

# Predicted values from the polynomial
y_pred = coefs[0] + coefs[1] * months + coefs[2] * months**2

# Residuals
residuals = values - y_pred

# Sum of squares of residuals
ss_res = np.sum(residuals**2)

# Total sum of squares
ss_tot = np.sum((values - np.mean(values))**2)

# R-squared
r_squared = 1 - (ss_res / ss_tot)
```

An (R2) value close to one (1) suggests that the model explains a large portion of the variance in the response variable, indicating a strong fit. Conversely, a value near zero (0) suggests the model does not explain the variance well.

## 2.4 Implementation

This section details the implementation of quadratic polynomial models in Python that are used in various applications, as demonstrated in this study. The core of the implementation involved the use of Python NumPy and Matplotlib libraries for mathematical operations and visualizations. The polynomial model is defined by the equation $ax2 + bx + c$, where $(a), (b), (c)$, are the coefficients optimized to fit the data points collected in different studies. The fitting process utilizes the `Polynomial.fit` method, which employs a least-squares polynomial fit. To ensure robustness and accuracy, the implementation also included the calculation of the coefficient of determination (R2) using NumPy's correlation function. This metric helps to assess the polynomial fit to the data, which is essential for the applications discussed, ranging from trend analysis in software usage to predicting the material properties of nanocomposites.

In the next step, we used Python to present an applied exploration of quadratic polynomial fitting and the coefficient of determination (R-squared) within the context of data analysis.

The following Python script is a practical implementation tool for researchers and analysts: It begins by prompting the user to describe the dataset, such as a location or a specific environmental metric, such as the PM2.5 air pollution index. This interactivity ensures that the resulting visualization is tailored and informative. The Python-formatted version of the code snippet of this part of our script is similar to:

```python
# User inputs for the descriptive elements of the plot

description = input("Enter the location description (e.g., Kyiv,
Shcherbakovskaya St.):")
pollution_name = input("Enter the pollution name (e.g., PM2.5):")
y_label = input("Enter the y-axis label (e.g., PM2.5 Index):")
```

The script reads data from a CSV file using Pandas, a library that excels in data manipulation. The data consists of monthly observations of the chosen metric. You can see this implementation in the Python-formatted version of the code snippet similar to:

```python
# Read data from a CSV file

# Use the direct link to the raw CSV file from the GitHub repository
data = pd.read_csv('https://raw.githubusercontent.com/rsipakov/QuadraticPolynomialsPyDA/main/notebooks/pm_data.csv')

# Or downloading CSV file to the local
# data = pd.read_csv('/path/pm_data.csv') # Update the path to your CSV file

months = data['Month'].to_numpy()
values = data['Values'].to_numpy()
```

As described above, with the data in hand, the NumPy library's `Polynomial.fit` function is employed to fit a quadratic polynomial to these observations. This is an essential step in modeling nonlinear behavior, accommodating potential fluctuations in data that a simple linear model would miss. Subsequently, the script computes the fitted values and leverages them to calculate the R-squared values. This statistic conveys the proportion of variance in the dependent variable explained by the independent variable. The Matplotlib library was then used to graphically represent the data along with the fitted curve, visually comparing the actual data points with those of the predictive model.

The following sources (VanderPlas, 2016; McKinney, 2017) provide a comprehensive overview of Python's theoretical background and practical application. These resources offer a deep dive into data analysis using Python, including comprehensive guidance on regression analysis, and robust examples that bridge theory with practice.

### 2.5. R-squared Limitations

R-squared is a commonly used metric for assessing the goodness of fit of regression models. However, it has notable limitations that must be taken into account when evaluating model performance. One key limitation is its sensitivity to the number of predictors in the model. As more predictors are added, R-squared will invariably increase, even if these additional predictors are not truly relevant to the model. This phenomenon can lead to overfitting, where the model captures noise rather than the actual relationship in the data (Fox & Weisberg, 2018).

To enhance the evaluation of regression models and address limitations such as overfitting, researchers can utilize a combination of metrics beyond the traditional R-squared. Adjusted

R-squared is recommended as it considers the number of predictors in the model, offering a more balanced assessment that considers model complexity and penalizes the inclusion of irrelevant predictors (Ajjaj et al., 2022). This adjustment helps prevent overfitting, ensuring that the model captures the underlying data patterns effectively. Additionally, Mean Squared Error (MSE) and Root Mean Squared Error (RMSE) are valuable metrics for assessing prediction accuracy by quantifying the average squared differences between observed and predicted values (Chicco et al., 2021). These metrics are particularly useful for comparing different models and providing an intuitive measure of prediction error.

To implement additional metrics like Adjusted R-squared, Mean Squared Error (MSE), and Root Mean Squared Error (RMSE) in our code, we provide below the Python-formatted version of the code snippet, similar to:

```python
# Calculate the Adjusted R-squared value

n = len(values)  # number of data points
p = 2  # number of predictors (polynomial degree)
adjusted_r_squared = 1 - ((1 - r_squared) * (n - 1) / (n - p - 1))

# Calculate Mean Squared Error (MSE) and Root Mean Squared Error (RMSE)

mse = mean_squared_error(values, y_pred)
rmse = np.sqrt(mse)
```

## 2.5 Operation

The software tool based on quadratic polynomial models requires the following system setup and workflow: Operating System—Windows, macOS, or Linux; Python Version—Python 3.6 or later; dependencies —NumPy, Matplotlib, Pandas, scikit-learn (the latest versions are recommended), memory, at least 4GB of RAM; Processor, minimum 1GHz processor, or faster. The software is accessible through https://mybinder.org, requires no local installation, and is fully configured to run in any web browser, ensuring its ease of use and reproducibility.

## 2.6 Installation process

To begin the installation process, it is imperative to ensure that Python is installed in the operating system. If Python is not present, it can be acquired from Python's official website https://python.org. After successful installation of Python, the next step involved installing the necessary libraries. This can be achieved through the Python Package Index (PyPI) using a PIP installer. Execute the following command in the command prompt or terminal to install the required libraries: `pip install numpy pandas matplotlib scikit-learn`. This command installs NumPy, which is essential for numerical computations, Matplotlib, a library for plotting graphs and effectively visualizing the data, Pandas, a data manipulation and analysis library, and scikit-learn is a widely used tool in the field of machine learning.

After developing the script using the quadratic polynomial models described above, the complete Python code was hosted on GitHub (Sipakov, 2024), enabling replication and further exploration of the findings. To facilitate ease of use and accessibility, the code was made available through MyBinder.org (https://mybinder.org/v2/gh/rsipakov/QuadraticPolynomialsPyDA/main), allowing it to operate in a live environment without the need for local setup. This implementation ensures that other researchers can directly interact with the codebase, providing a dynamic way to validate and extend research findings.

## 3. Results

The quadratic polynomial fitting method used in this study demonstrates Python's ability to effectively manage and analyze complex datasets. The datasets used herein are illustratively generated, serving as a basis for demonstrating the potential applications of quadratic polynomial models. The fitting process provided a smooth curve aligned closely with the observed data points, indicating robust model performance. Notably, the computed coefficient of determination, R-squared ($R^2$), was substantially high, reflecting a strong correlation between the observed values and those predicted by the model. This statistical measure underpins the polynomial's ability to capture and explain variability in the data effectively, which is crucial for validating the regression model used in this analysis. Figure 2 illustrates the quadratic polynomial curve fitted to the observed data points using Python's plotting library Matplotlib.

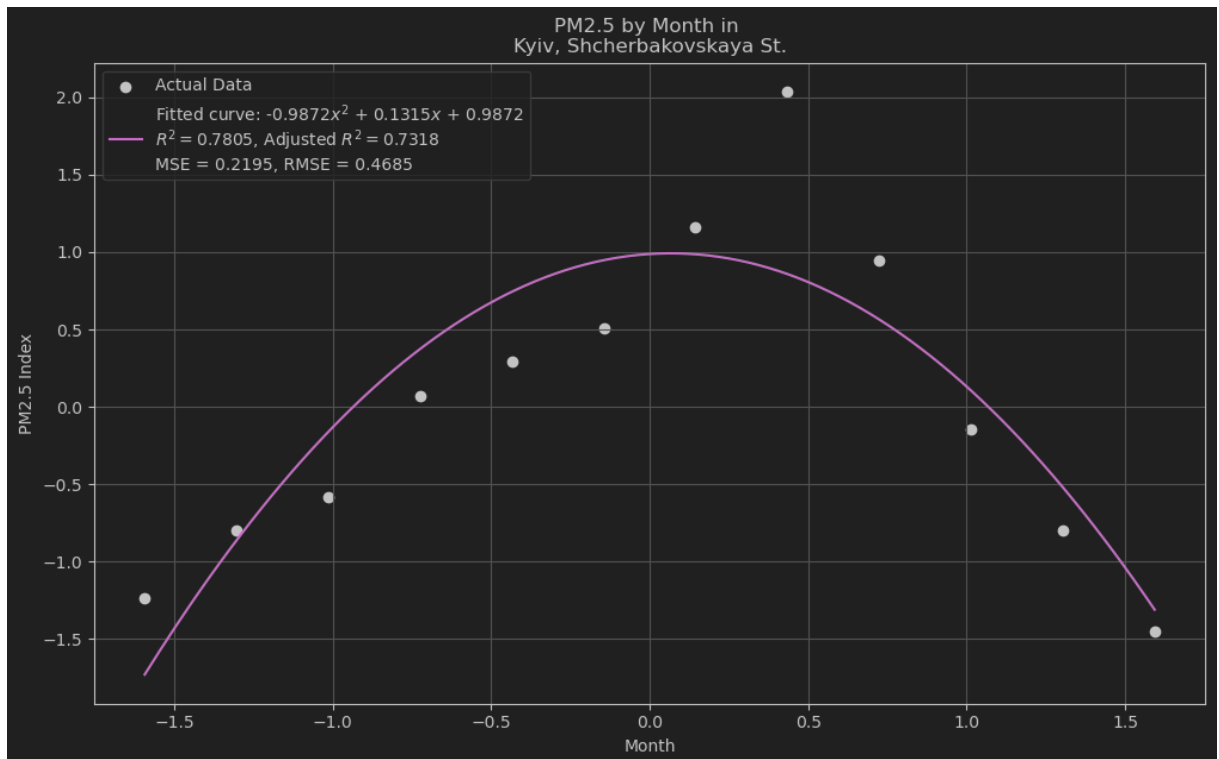

Figure 2. The quadratic polynomial fit of the dataset.

The curve represents the model obtained from regression analysis, where the quadratic polynomial provides a significant fit to the data, as evidenced by the computed R-squared value. The axes were labeled to identify the independent variable (x-axis) and dependent variable (y-axis), and a legend was included to differentiate between the observed data points and fitted polynomial curve. The smoothness of the curve indicates the effectiveness of the model in capturing trends within the dataset, which can be utilized for predictive analytics and further statistical inferences. After configuring the plot with the necessary parameters for clear and informative visualization, it was generated using the `plt.show()` function in Matplotlib.

To provide a broader perspective on quadratic models' performance, we compared them with linear and cubic models using the same datasets, and the results as shown in Figure 3. This comparison utilized three key metrics: R-squared, Adjusted R-squared, and Mean Squared Error (MSE) to assess the fit and predictive accuracy. We used Python to fit linear, quadratic, and cubic models to the datasets and calculate the aforementioned metrics. The results of the model fitting are summarized in the Table.

Table 1. Comparing quadratic, linear, and cubic models using the same datasets

| Model | R-squared | Adjusted R-squared | MSE |
|---|---|---|---|
| Linear | 0.0173 | -0.0810 | 0.9827 |
| Quadratic | 0.7805 | 0.7318 | 0.2195 |
| Cubic | 0.8737 | 0.8263 | 0.1263 |

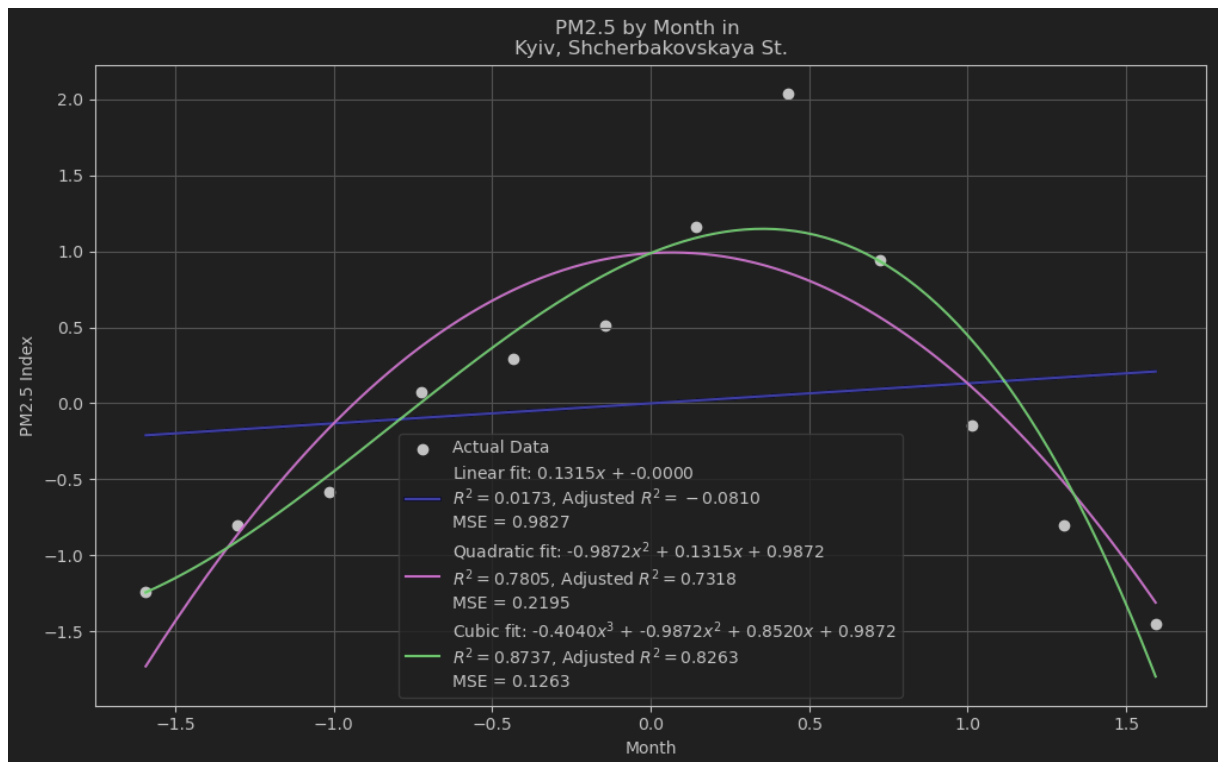

Figure 3. The quadratic polynomial fit of the dataset. Comparing quadratic, linear, and cubic models using the same datasets

The linear model shows a very poor fit with an R-squared value of 0.0173, indicating that it barely explains the variance in the data. The negative Adjusted R-squared suggests that the model is not suitable for this dataset. The quadratic model significantly improves the fit, with an R-squared value of 0.7805 and an Adjusted R-squared of 0.7318, indicating a much better fit compared to the linear model. The MSE is also substantially lower. The cubic model provides the best fit among the three, with the highest R-squared value of 0.8737 and Adjusted R-squared of 0.8263, along with the lowest MSE, indicating that it explains the variance in the data the most effectively.

The selection between quadratic and cubic models should account for the intricacy of the dataset and the potential for overfitting. The quadratic model often strikes an optimal balance between simplicity and precision, making it preferable for numerous practical applications. The quadratic model's ability to effectively capture the essential patterns in the data without the risk of overfitting associated with higher-degree polynomials underscores its utility and robustness in empirical research and applied data analysis.

## 4. Discussion

Quadratic polynomials are valued for their ability to model nonlinear relationships in various data contexts, balancing computational efficiency, and interpretability. However, their performance can be limited when confronted with complex multivariable systems in which more sophisticated statistical models may be more accurate. Future research could address these challenges by focusing on several advancements in quadratic polynomial modeling. Incorporating regularization techniques is recommended to counteract overfitting, particularly for datasets with intricate structures. Exploring hybrid models that merge the clear interpretive benefits of quadratic polynomials with the robust capabilities of machine learning algorithms could also enhance predictive accuracy.

Moreover, the development of adaptive polynomial models that adjust their parameters based on real-time data inputs can significantly improve the dynamic data analysis. Extending these models to operate within multiscale frameworks may offer deeper insights into various levels of data structure, ensuring a comprehensive understanding of complex patterns. These enhancements are crucial for extending the utility of quadratic polynomials beyond their current capabilities and facilitating more accurate and efficient statistical analyses across diverse datasets.

This study acknowledges that the effectiveness of quadratic polynomials, like any statistical model, is contingent on the quality and volume of the data available. To mitigate potential biases and inaccuracies in the input data, the data collection methodology should include rigorous data preprocessing steps, such as outlier removal, normalization, and feature selection, which are crucial for enhancing the reliability of the research. Despite the potential of more advanced models, this study primarily advocates quadratic polynomials because of their suitability for datasets exhibiting quadratic relationships, which are frequently encountered in environment-related target research. However, future research should continue to explore the comparative dynamic performance of quadratic polynomials, for example, the performance of benchmarking against contemporary machine-learning algorithms to ensure a comprehensive understanding of their relative merits, possibly extending the use of hybrid approaches that combine the strengths of traditional polynomial models and cutting-edge machine-learning techniques.

While quadratic models offer simplicity and clarity, they may only capture part of the complexity of data as effectively as some machine-learning models. However, their computational efficiency and suitability for smaller datasets can be advantageous for specific scenarios.

**Challenges and Limitations of Quadratic Polynomials**

Quadratic polynomials can effectively model nonlinear relationships, but they are prone to overfitting, particularly with small or noisy datasets. Overfitting occurs when the model becomes too complex and captures the noise in the data rather than the underlying trend. This results in poor generalization to new data.

When the dataset is small, the quadratic model might fit the few available data points too closely, capturing random variations rather than the actual trend. In datasets with a high level of noise, a quadratic polynomial might try to model these fluctuations, leading to overfitting.

Techniques such as Ridge Regression (L2 regularization) or Lasso Regression (L1 regularization) can penalize the complexity of the model, discouraging overfitting. Using cross-validation methods, such as k-fold cross-validation, helps in assessing the model's performance on different subsets of the data, ensuring it generalizes well to unseen data. Sometimes, a linear model might suffice if the relationship is not strongly nonlinear. Alternatively, we can consider higher-order polynomials only if there is a clear justification from the data.

Quadratic models are sensitive to variability in the data, which can significantly affect their performance. Small changes in the data can lead to substantial changes in the fitted model, impacting its stability and predictive power. Properly preprocessing the data by removing outliers and handling missing values can improve model stability. Implementing robust statistical techniques to detect and handle anomalies can mitigate the sensitivity of the model to data variability. Combining multiple models (e.g., through bagging or boosting) can reduce the impact of variability and enhance predictive performance.

**When Regularization and Cross-Validation are Needed**

Unfortunately, `Polynomial.fit` itself does not support regularization directly, so we need to mimic regularization and cross-validation manually. However, the use of `Polynomial.fit` from the NumPy library for fitting a polynomial to data has its advantages, particularly for straightforward polynomial regression without the need for additional regularization or cross-validation. The syntax is simple and easy to understand, making the code more readable. `Polynomial.fit` directly provides the coefficients of the fitted polynomial, which can be easily accessed and interpreted. For small to moderate-sized datasets, `Polynomial.fit` is computationally efficient. It uses least-squares fitting, which is optimal for such scenarios without the overhead of more complex methods. This can be particularly useful when the primary goal is to quickly fit and visualize a polynomial model.

While `Polynomial.fit` has these advantages, there are situations where more complex techniques like regularization and cross-validation become necessary. `Polynomial.fit` is excellent for simplicity, quick fitting, and visualization. Ridge regression with cross-validation offers more robust performance evaluation and overfitting prevention for complex scenarios.

## 5. Conclusion

In this study, we demonstrate the practical utility and versatility of quadratic polynomials for advanced data analysis in Python. We showcase how quadratic models can effectively capture nonlinear relationships in complex datasets by leveraging the capabilities of Python libraries such as NumPy, Matplotlib, scikit-learn, and Pandas.

The results highlight the strong performance of the quadratic polynomial fit, as evidenced by the high coefficient of determination (R-squared) value. This indicates that the model explains a substantial proportion of the variability in the data. Comparisons with linear and cubic models further underscore the quadratic model's ability to strike an optimal balance between simplicity and precision for many practical applications.

However, our study also acknowledges quadratic polynomials' limitations, such as their susceptibility to overfitting with small or noisy datasets and their sensitivity to data variability. Techniques like regularization and cross-validation are recommended to mitigate these challenges and ensure robust model performance.

Future research directions are proposed, including the incorporation of regularization methods, exploration of hybrid models combining quadratic polynomials with machine learning algorithms, development of adaptive polynomial models, and extension to multiscale frameworks. These advancements can further enhance the accuracy and efficiency of quadratic polynomial models for diverse data analysis tasks.

## Ethical compliance

All procedures involving human participants were performed in accordance with the ethical standards of the Institutional and National Research Committee.

## Author contributions

Rostyslav Sipakov contributed to the research design, implementation, and manuscript writing. Dr. Voloshkina and Dr. Kovalova helped implement and analyze the results. All authors have seen and agreed to the final content of the manuscript.

## Data availability statement

No data is associated with this article.

## Software availability statement

- Source code of the scripts available from: https://github.com/rsipakov/QuadraticPolynomialsPyDA
- Archived scripts available from: https://doi.org/10.5281/zenodo.10637508
- License: OSI approved open license software is under MIT License (https://opensource.org/license/MIT)

## Previous version

Sipakov R, Voloshkina O and Kovalova A. Leveraging Quadratic Polynomials in Python for Advanced Data Analysis [version 1; peer review: 1 approved with reservations]. *F1000Research* 2024, **13**:490 (https://doi.org/10.12688/f1000research.149391.1)

## Reviewer report

Molla S. Peer Review Report For: Leveraging Quadratic Polynomials in Python for Advanced Data Analysis [version 1; peer review: 1 approved with reservations]. *F1000Research* 2024, **13**:490 (https://doi.org/10.5256/f1000research.163848.r305465)